\documentclass[aps,amsmath,amssymb,superscriptaddress,prb,reprint,twocolumn]{revtex4-1}
\usepackage{graphicx}% Include figure files
\usepackage{dcolumn}% Align table columns on decimal point
\usepackage{braket,bm}% bold math
\usepackage[utf8]{inputenc}
\usepackage[T1]{fontenc}
\usepackage{mathptmx}

\begin{document}

\title{Optical induced Spin Current in Monolayer NbSe$_2$}

\author{Ren Habara}
\affiliation{Department of Nanotechnology for Sustainable Energy, School
of Science and Technology, Kwansei Gakuin University, Gakuen 2-1, Sanda 669-1337, Japan}
\author{Katsunori Wakabayashi}
\affiliation{Department of Nanotechnology for Sustainable Energy, School
of Science and Technology, Kwansei Gakuin University, Gakuen 2-1, Sanda 669-1337, Japan}
\affiliation{National Institute for Materials Science (NIMS), Namiki 1-1, Tsukuba 305-0044, Japan}
\affiliation{Center for Spintronics Research Network (CSRN), Osaka
University, Toyonaka 560-8531, Japan}

\date{\today}

\begin{abstract}
Monolayer NbSe$_2$ is a metallic two-dimensional (2D) transition-metal
 dichalcogenide material. Owing to the lattice structure and the strong
 atomic spin-orbit coupling (SOC) 
 field, monolayer NbSe$_2$ possesses Ising-type SOC which acts as
 effective Zeeman field, leading to the unconventional 
 topological spin properties. In this paper, we numerically calculate
 spin-dependent optical conductivity of monolayer NbSe$_2$ using Kubo
 formula based on an effective tight-binding model which includes
 $d_{z^2}$, $d_{x^2-y^2}$ and $d_{xy}$ orbitals of Nb atom. 
 Numerical calculation indicates that the up-
 and down-spin have opposite sign of Hall current, so the pure spin Hall
 current can be generated in monolayer NbSe$_2$ under light irradiation, owing
 to the topological nature of monolayer NbSe$_2$, i.e., finite spin Berry curvature.
 The spin Hall angle is also evaluated.
 The optical induced spin Hall current can be enhanced by the electron
 doping and persists even at room temperature. 
 Our results will serve to design opt-spintronics devices such as {\it
 spin current harvesting by light irradiation}
 on the basis of 2D materials.
\end{abstract}

\maketitle

Transition-metal dichalcogenide (TMDC) is a new class of two-dimensional
(2D) electronic systems and provides a platform to design new
functional opt-electronic devices.~\cite{Mak2010, Sple2010, Ton2012,
Gut2013, Zhao2013} In TMDC, electronic properties
crucially depend on the combination of metal and chalcogen
atoms.~\cite{Wang2012, Chhowalla2013}
Because of weak van der Waals forces between layers,
monolayer of TMDC can be easily exfoliated and exhibit many fascinating
properties such as valley-dependent optical selection
rule~\cite{Zeng, Mak, Cao, Yu} and spin Hall
effect (SHE).~\cite{Sinova, Hai, Kato, Wun}
In particular, monolayer NbSe$_2$ belongs to monolayer of group-$\rm{V}$ TMDC MX$_2$ (M=Nb, Ta; X=S, Se), and is known to show
metallic behavior with superconducting phase transition at low
temperature.~\cite{Kim, He, Xi, Sohn}
Monolayer NbSe$_2$ has a hexagonal lattice structure but with no spatial inversion
symmetry and possesses out-of-plane mirror symmetry. 
Owing to the lattice structure and strong atomic spin-orbit coupling (SOC) field, 
monolayer NbSe$_2$ possesses Ising-type
SOC,~\cite{He, Xi, Sohn, Lu, Saito, Zhou, Baw} i.e., effective Zeeman field that locks electron spin to
out-of-plane directions by in-plane momentum
and provides the unconventional topological spin properties.

\begin{figure}[h]
  \begin{center}
    \includegraphics[width=0.42\textwidth]{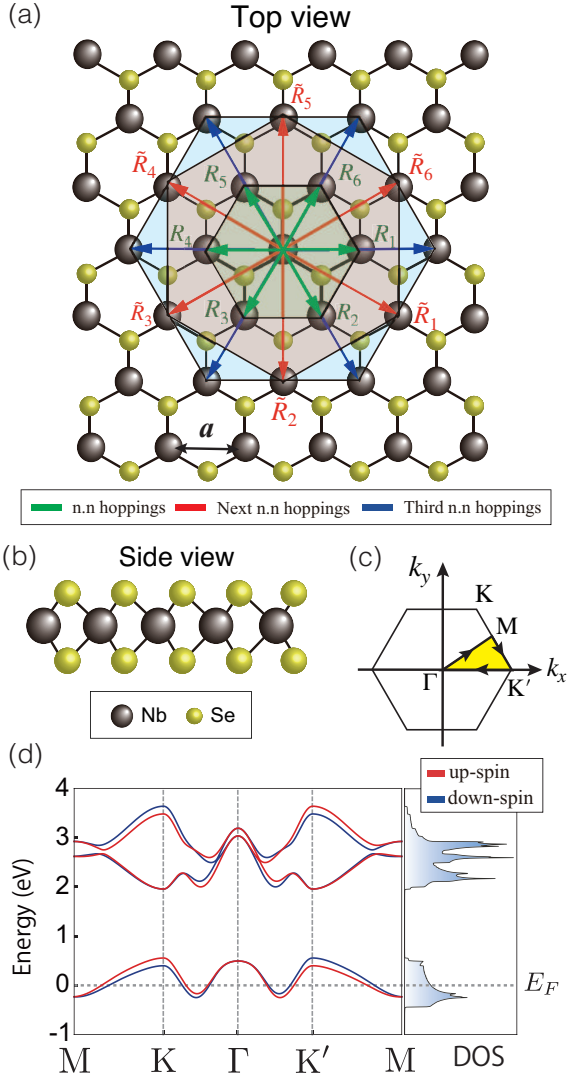}
   \caption{Crystal structure of monolayer NbSe$_2$ which consists of Nb
   (black) and Se (yellow) atoms. (a) Top view and (b) side view of the
   lattice structure. Green, red and blue arrows indicate hopping vectors
   $\bm{R_i}$ $(i=1,2,\cdots,6)$ pointing to n.n sites,
   the vectors $\bm{\tilde{R_j}}$ $(j=1,2,\cdots,6)$ pointing to next n.n
   sites and the vectors $2\bm{R_i}$ pointing to third n.n sites, respectively. $a$ is the
   lattice constant. (c) 1st BZ of monolayer NbSe$_2$. (d)
   Energy band structure and DOS of NbSe$_2$
   with SOC parameter $\lambda_{SOC}=0.0784$ eV. Fermi level is set to
   zero.}
    \label{fig:1}
  \end{center}
\end{figure}

In this paper, we show that pure spin Hall current can be generated in
monolayer NbSe$_2$ by light irradiation, owing to the topological nature
of monolayer NbSe$_2$, i.e., finite spin Berry curvature. 
Figures~\ref{fig:1} (a) and (b) show the top and side views of monolayer
NbSe$_2$, respectively, where a layer of Nb atoms is sandwiched by two layers of Se
atoms. 
From top view, monolayer NbSe$_2$ has a hexagonal lattice structure but with no spatial inversion
symmetry. Also, from the side view, it respects out-of-plane mirror
symmetry. 
Figure~\ref{fig:1} (c) shows the corresponding first Brillouin Zone (BZ). 
We employ a multi-orbitals tight-binding model (TBM) which includes $d_{z^2}$, $d_{x^2-y^2}$
and $d_{xy}$ orbitals of Nb atom to describe the electronic states of
NbSe$_2$.~\cite{He, Liu} 
The eigenvalue equation for TBM is 
\begin{math}
 \hat{H}(\bm{k})|u_{n\bm{k}}\rangle = E_{n\bm{k}}|u_{n\bm{k}}\rangle,
\end{math}
where $\bm{k}=(k_x,k_y)$ is the wavenumber vector, $E_{n\bm{k}}$ is the eigenvalue and $n=1,2,\cdots,6$
is the band index. 
The eigenvector is defined as 
$|u_{n\bm{k}}\rangle =
(c_{n\bm{k},d_{z^2},\uparrow},c_{n\bm{k},d_{xy},\uparrow},c_{n\bm{k},d_{x^2-y^2},\uparrow},c_{n\bm{k},d_{z^2},\downarrow},c_{n\bm{k},d_{xy},\downarrow},c_{n\bm{k},d_{x^2-y^2},\downarrow})^T$, 
where $(\cdots)^T$ indicates the transpose of vector and
$c_{n\bm{k}\tau s}$ 
means the amplitude at atomic orbital $\tau$ with spin $s$ for the $n$th energy band at $\bm{k}$.
The Hamiltonian with the SOC can be written as
\begin{equation}
 \hat{H}(\bm{k})=\hat{\sigma}_0\otimes \hat{H}_{TNN}(\bm{k})+\hat{\sigma}_z\otimes\frac{1}{2}\lambda_{SOC} \hat{L}_z
\end{equation}
with
\begin{equation}
 \hat{H}_{TNN}(\bm{k})=
  \begin{pmatrix}
   V_{0}&V_{1}&V_{2}\\
   V_{1}^{*}&V_{11}&V_{12}\\
   V_{2}^{*}&V_{12}^{*}&V_{22}\\
  \end{pmatrix}
\end{equation}
and
\begin{equation}
 \hat{L}_z=
  \begin{pmatrix}
   0&0&0\\
   0&0&-2i\\
   0&2i&0\\
  \end{pmatrix}.
\end{equation}
Here, $\hat{\sigma}_0$ and $\hat{\sigma}_z$ are Pauli matrices, and
$\lambda_{SOC}$ is the Ising-type SOC parameter. In monolayer NbSe$_2$,
$\lambda_{SOC}=0.0784$ eV.
$\hat{H}_{TNN}(\bm{k})$ includes the electron hoppings only among three
$d$-orbitals of Nb atoms, which are assumed up to third-nearest neighbor
sites as shown in Fig.~\ref{fig:1} (a). 
Here, green, red and blue arrows
indicate hopping vectors $\bm{R_i}$ $(i=1,2,\cdots,6)$
pointing to nearest-neighbor (n.n) sites, the vectors $\bm{\tilde{R_j}}$
$(j=1,2,\cdots,6)$ pointing to next n.n sites and the vectors $2\bm{R_i}$
pointing to third n.n sites, respectively. 
The details of matrix elements $V_0$, $V_1$, $V_2$, $V_{11}$, $V_{12}$ and $V_{22}$
can be found in Supplementary material.~\cite{supplementary}
Figure~\ref{fig:1} (d) shows the energy band structure of monolayer
NbSe$_2$ along the line passing through the high-symmetric points of 1st
BZ and the corresponding density of states (DOS). Here, red and blue lines indicate spin-up and spin-down
states. Monolayer NbSe$_2$ is metallic, but a large 
energy band gap between the partially filled valence bands and empty
conduction bands. Also, the Ising-type SOC provides the opposite spin
splitting at the valence band edges in K and K$^{\prime}$ points, and time-reversal symmetry protection.

We numerically calculate spin-dependent optical conductivity of
NbSe$_2$ using Kubo formula~\cite{Qiao, Guo, Seng, Varg, Tanaka,
Ferreira, Morimoto, Yugui, LiZ, Akita} based on an effective TBM, and find the photo-induced generation of pure
spin Hall current.~\cite{Huang, Lin, ShanW} The spin-dependent optical Hall conductivity can be given as 
\begin{equation}
 \begin{split}
  \sigma^{spin}_{xy}(\omega) & =
  \frac{i{\hbar}e}{S}\sum_{\bm{k}}\sum_{nm}\frac{f(E_{n\bm{k}})-f(E_{m\bm{k}})}{E_{m\bm{k}}-E_{n\bm{k}}} \\
& \cdot\frac{\braket{u_{n\bm{k}}|\hat{j}^{spin}_x|u_{m\bm{k}}}\braket{u_{m\bm{k}}|\hat{v}_y|u_{n\bm{k}}}}{E_{m\bm{k}}-E_{n\bm{k}}-\hbar\omega-i\eta}, 
 \end{split}
\label{eq:kubo}
\end{equation}
where $n$($m$) indicates the band index including spin degree
of freedom, $\Ket{u_{n\bm{k}}}$ is the eigen function with the
eigen energy $E_{n\bm{k}}$ and $f(E_{n\bm{k}})$ is Fermi-Dirac
distribution function. $\eta$ is infinitesimally small
real number, and $S$ is the area of system.
Also, $\hat{j}^{spin}_x$ is the spin current operator, and written as
$\hat{j}^{spin}_x=\frac{1}{2}\{\frac{\hbar}{2}\hat{\sigma}_z\otimes \hat{I}_3, \hat{v}_x\}$,
where $\hat{I}_3$ is the $3\times 3$ identity matrix and
$(\hat{v}_{x},\hat{v}_{y})=\frac{1}{\hbar}(\frac{\partial
\hat{H}}{\partial x},\frac{\partial \hat{H}}{\partial y})$ is the group
velocity operator. We add the superscript $spin$ for the
spin-dependent optical Hall conductivity in order to distinguish its
conductivity from ordinary optical Hall conductivity without SOC. When we
consider $\sigma^{spin}_{xy}(\omega)$ for direct current (DC) limit ($\omega=0$),
zero-temperature ($T=0$ K) and clean limit ($\eta=0$ eV), Equation
(14) becomes
\begin{equation}
 \sigma^{spin}_{xy}=\frac{e}{S}\sum_{\bm{k}}\Omega^{spin}(\bm{k})
\label{eq.spinhallconductivity}
\end{equation}
with
\begin{equation}
  \Omega^{spin}(\bm{k})=\hbar\sum_{n}f(E_{n\bm{k}})\sum_{m\neq n}\frac{-2\mathrm{Im}\braket{u_{n\bm{k}}|\hat{j}^{spin}_x|u_{m\bm{k}}}\braket{u_{m\bm{k}}|\hat{v}_y|u_{n\bm{k}}}}{(E_{m\bm{k}}-E_{n\bm{k}})^2}.
\label{eq:spinBerrycurvature}
\end{equation}
Here, $\Omega^{spin}(\bm{k})$ is spin Berry curvature of Bloch
state,~\cite{Qiao, Slawinska, Yao, Da, Grad, Qu, Ki, GuoG, Yugui, Zh}
and drives an anomalous transverse velocity~\cite{Feng, Xiao} written as
\begin{equation}
 v_{\perp}=-\frac{e}{\hbar}\bm{E}\times\Omega^{spin}(\bm{k})
\end{equation}
under the presence of an electric field $\bm{E}$. 

\begin{figure}[h]
  \begin{center}
    \includegraphics[width=0.372\textwidth]{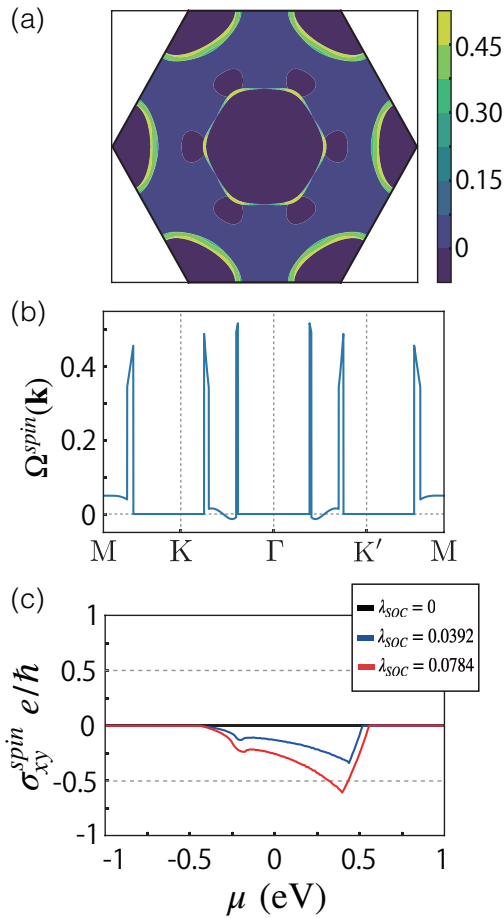}
    \caption{Spin Berry curvature of monolayer NbSe$_2$ with SOC
   parameter $\lambda_{SOC}=0.0784$ eV. (a) Contour plot in 1st
   BZ. (b) Plot along the path through the high-symmetric points in 1st
   BZ. (c) Fermi-level dependence of spin-dependent Hall conductivity
   $\sigma^{spin}_{xy}$ of DC limit for several different SOC parameters. The
   unit of $\sigma^{spin}_{xy}$ is $e^2/\hbar$.}
    \label{fig:2}
  \end{center}
\end{figure}
Figure~\ref{fig:2} (a) shows the 2D contour plot of spin Berry curvature in
1st BZ. It clearly shows the six-fold symmetry. However, for each spin
state, its symmetry around $\Gamma$ point reduces to three-fold symmetry
because the energy band of monolayer NbSe$_2$ has opposite spin
splitting around K and K$^{\prime}$ points [see Figs.~S2 (a) and (b)].~\cite{supplementary} In Fig.~2 (b), the spin Berry curvature is plotted 
 along the path passing through high-symmetric points of 1st BZ.
Owing to the Fermi-Dirac distribution function in Eq.~(6), the spin Berry
curvature is vanished at valleys.
This is very contrast from semiconducting MX$_2$ such as MoS$_2$, where
valley current is induced owing to the finite Berry curvature at the valleys.
In monolayer NbSe$_2$, valley current can be induced by the electron
doping [see Figs.~S4 (a) and (b)].~\cite{supplementary}
As can be seen in Fig.~\ref{fig:2} (b), the spin Berry curvature has sharp peaks with the
same sign, because the energy differences between up and down spin states become
very small at the Fermi energy as can be confirmed in Fig.~\ref{fig:1} (d). 
Figure~\ref{fig:2} (c) shows Fermi energy dependence of spin-dependent Hall conductivity for several different SOC
parameters. Since spin Hall conductivity is the summation of the spin Berry
curvature over the momentum in DC limit, and can be calculated by using
Eq.~(\ref{eq.spinhallconductivity}). 
It is noted that the spin Hall conductivity is enhanced by electron-doping.
Thus, finite Berry curvature manifests the intrinsic SHE~\cite{Zh, GuoG,
Feng}, i.e.,
the generation of transverse spin current by the application of electric
field. 
%In further, by irradiating light
%in monolayer NbSe$_2$, we find optical induced transverse spin current.

\begin{figure}[h]
  \begin{center}
    \includegraphics[width=0.48\textwidth]{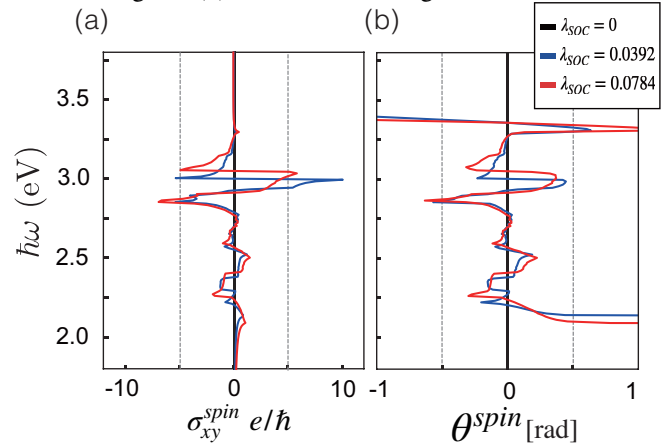}
    \caption{Optical angular frequency dependence of (a) spin-dependent Hall
   conductivity $\sigma^{spin}_{xy}(\omega)$ of $E_F=0$ eV and (b) SHA $\theta^{spin}$ for several different SOC parameters. The unit of
   $\sigma^{spin}_{xy}(\omega)$ is $e^2/\hbar$.}
    \label{fig:3}
  \end{center}
\end{figure}
Figure~\ref{fig:3} (a) shows the numerically calculated angular frequency
dependence of spin-dependent optical Hall conductivity for several
different SOC parameters. 
%Here, the unit of the spin-dependent optical
%Hall conductivity is $e^2/\hbar$, and we set $E_F=0$ eV. 
It is clearly seen that the large peak around $3.0$ eV, i.e., the generation
of spin Hall current in NbSe$_2$ ($\lambda_{SOC}=0.0784$ eV) by
light irradiation. 
The cases for $\lambda_{SOC}=0.0392$ and $0$ eV are also plotted for the
comparison.
It should be noted that  
the optical charge Hall conductivity is identically zero 
for arbitrary energy, because the charge Berry curvature has the
anti-symmetric with the wave number, i.e., 
$\Omega^{charge}(\bm{k})=-\Omega^{charge}(-\bm{k})$ [see
Figs.~S2 (c) and (d)].~\cite{supplementary} 
Thus, the pure spin current can be induced by light irradiation in
monolayer NbSe$_2$. 
%It should be noted that  
%the spin Hall current disappears in the limit of $\lambda_{SOC}=0$ eV.
Since the intensive peaks of $\sigma^{spin}_{xy}$ appear in the range of
$2.7-3.2$ eV, i.e., visible and ultraviolet range, 
monolayer NbSe$_2$ can be used for the application of {\it spin current
harvesting by light irradiation}. 
The effect of optical induced spin Hall current is also robust to
temperature, and is expected to be observed even in room temperature [see
Fig.~S5 (a)].~\cite{supplementary}

Figure~3 (b) shows the angular frequency dependence of spin Hall angle (SHA),~\cite{Qiao, Zh,
Huang, WangY, ZhanW} which measures conversion efficiency from charge current to spin
current. SHA is given as
\begin{equation}
 \theta^{spin}=\frac{2e}{\hbar}\frac{\sigma^{spin}_{xy}}{\sigma_{xx}},
\end{equation}
where $\sigma_{xx}$ is optical longitudinal conductivity~\cite{Kubo,
Ferreira, SPS} which is found in Fig.~S5 (b).~\cite{supplementary} 
SHA has divergence below about $2.1$ eV and above about $3.3$ eV, 
because $\sigma_{xx}$ becomes zero, where SHA becomes ill-defined.
When the Hall conductivity has peak, the SHA also has larger value and then
indicates that we can generate the pure spin current efficiently.
Moreover, the observed optical induced pure Hall spin current is robust
against the carrier doping. In particular, the pure spin current can be
enhanced by electron-doping [see Fig.~S6].~\cite{supplementary} These results
indicate that we can generate the spin Hall current efficiently by light
irradiation using visible light of violet 
and enhance its magnitude by electron doping.

In conclusion, we have theoretically proposed that pure spin Hall current
can be induced efficiently in monolayer NbSe$_2$ 
by irradiating the visible light. The origin of spin Hall current can be
attributed by the finite spin Berry curvature owing to the Ising-type SOC in
monolayer NbSe$_2$.
It is also found that the optical induced spin Hall current can be enhanced by the
electron doping and persists even in the room temperature[see
Supplementary material].~\cite{supplementary}
Thus, monolayer NbSe$_2$ can be used for the source of pure spin current
by using light irradiation. 
 Our results will serve to design opt-spintronics devices such as {\it
 spin current harvesting by light irradiation}
 on the basis of 2D materials.

This work was supported by JSPS KAKENHI
(Nos. JP21H01019, JP18H01154) and JST CREST (No. JPMJCR19T1).

\nocite{*}
\bibliography{reference}

\begin{thebibliography}{52}%
\makeatletter
\providecommand \@ifxundefined [1]{%
 \@ifx{#1\undefined}
}%
\providecommand \@ifnum [1]{%
 \ifnum #1\expandafter \@firstoftwo
 \else \expandafter \@secondoftwo
 \fi
}%
\providecommand \@ifx [1]{%
 \ifx #1\expandafter \@firstoftwo
 \else \expandafter \@secondoftwo
 \fi
}%
\providecommand \natexlab [1]{#1}%
\providecommand \enquote  [1]{``#1''}%
\providecommand \bibnamefont  [1]{#1}%
\providecommand \bibfnamefont [1]{#1}%
\providecommand \citenamefont [1]{#1}%
\providecommand \href@noop [0]{\@secondoftwo}%
\providecommand \href [0]{\begingroup \@sanitize@url \@href}%
\providecommand \@href[1]{\@@startlink{#1}\@@href}%
\providecommand \@@href[1]{\endgroup#1\@@endlink}%
\providecommand \@sanitize@url [0]{\catcode `\\12\catcode `\$12\catcode
  `\&12\catcode `\#12\catcode `\^12\catcode `\_12\catcode `\%12\relax}%
\providecommand \@@startlink[1]{}%
\providecommand \@@endlink[0]{}%
\providecommand \url  [0]{\begingroup\@sanitize@url \@url }%
\providecommand \@url [1]{\endgroup\@href {#1}{\urlprefix }}%
\providecommand \urlprefix  [0]{URL }%
\providecommand \Eprint [0]{\href }%
\providecommand \doibase [0]{http://dx.doi.org/}%
\providecommand \selectlanguage [0]{\@gobble}%
\providecommand \bibinfo  [0]{\@secondoftwo}%
\providecommand \bibfield  [0]{\@secondoftwo}%
\providecommand \translation [1]{[#1]}%
\providecommand \BibitemOpen [0]{}%
\providecommand \bibitemStop [0]{}%
\providecommand \bibitemNoStop [0]{.\EOS\space}%
\providecommand \EOS [0]{\spacefactor3000\relax}%
\providecommand \BibitemShut  [1]{\csname bibitem#1\endcsname}%
\let\auto@bib@innerbib\@empty
%</preamble>
\bibitem [{\citenamefont {Mak}\ \emph {et~al.}(2010)\citenamefont {Mak},
  \citenamefont {Lee}, \citenamefont {Hone}, \citenamefont {Shan},\ and\
  \citenamefont {Heinz}}]{Mak2010}%
  \BibitemOpen
  \bibfield  {author} {\bibinfo {author} {\bibfnamefont {K.~F.}\ \bibnamefont
  {Mak}}, \bibinfo {author} {\bibfnamefont {C.}~\bibnamefont {Lee}}, \bibinfo
  {author} {\bibfnamefont {J.}~\bibnamefont {Hone}}, \bibinfo {author}
  {\bibfnamefont {J.}~\bibnamefont {Shan}}, \ and\ \bibinfo {author}
  {\bibfnamefont {T.~F.}\ \bibnamefont {Heinz}},\ }\href {\doibase
  10.1103/physrevlett.105.136805} {\bibfield  {journal} {\bibinfo  {journal}
  {Phys. Rev. Lett.}\ }\textbf {\bibinfo {volume} {105}},\ \bibinfo {pages}
  {136805} (\bibinfo {year} {2010})}\BibitemShut {NoStop}%
\bibitem [{\citenamefont {Splendiani}\ \emph {et~al.}(2010)\citenamefont
  {Splendiani}, \citenamefont {Sun}, \citenamefont {Zhang}, \citenamefont {Li},
  \citenamefont {Kim}, \citenamefont {Chim}, \citenamefont {Galli},\ and\
  \citenamefont {Wang}}]{Sple2010}%
  \BibitemOpen
  \bibfield  {author} {\bibinfo {author} {\bibfnamefont {A.}~\bibnamefont
  {Splendiani}}, \bibinfo {author} {\bibfnamefont {L.}~\bibnamefont {Sun}},
  \bibinfo {author} {\bibfnamefont {Y.}~\bibnamefont {Zhang}}, \bibinfo
  {author} {\bibfnamefont {T.}~\bibnamefont {Li}}, \bibinfo {author}
  {\bibfnamefont {J.}~\bibnamefont {Kim}}, \bibinfo {author} {\bibfnamefont
  {C.-Y.}\ \bibnamefont {Chim}}, \bibinfo {author} {\bibfnamefont
  {G.}~\bibnamefont {Galli}}, \ and\ \bibinfo {author} {\bibfnamefont
  {F.}~\bibnamefont {Wang}},\ }\href {\doibase 10.1021/nl903868w} {\bibfield
  {journal} {\bibinfo  {journal} {Nano Lett.}\ }\textbf {\bibinfo {volume}
  {10}},\ \bibinfo {pages} {1271} (\bibinfo {year} {2010})}\BibitemShut
  {NoStop}%
\bibitem [{\citenamefont {Tongay}\ \emph {et~al.}(2012)\citenamefont {Tongay},
  \citenamefont {Zhou}, \citenamefont {Ataca}, \citenamefont {Lo},
  \citenamefont {Matthews}, \citenamefont {Li}, \citenamefont {Grossman},\ and\
  \citenamefont {Wu}}]{Ton2012}%
  \BibitemOpen
  \bibfield  {author} {\bibinfo {author} {\bibfnamefont {S.}~\bibnamefont
  {Tongay}}, \bibinfo {author} {\bibfnamefont {J.}~\bibnamefont {Zhou}},
  \bibinfo {author} {\bibfnamefont {C.}~\bibnamefont {Ataca}}, \bibinfo
  {author} {\bibfnamefont {K.}~\bibnamefont {Lo}}, \bibinfo {author}
  {\bibfnamefont {T.~S.}\ \bibnamefont {Matthews}}, \bibinfo {author}
  {\bibfnamefont {J.}~\bibnamefont {Li}}, \bibinfo {author} {\bibfnamefont
  {J.~C.}\ \bibnamefont {Grossman}}, \ and\ \bibinfo {author} {\bibfnamefont
  {J.}~\bibnamefont {Wu}},\ }\href {\doibase 10.1021/nl302584w} {\bibfield
  {journal} {\bibinfo  {journal} {Nano Lett.}\ }\textbf {\bibinfo {volume}
  {12}},\ \bibinfo {pages} {5576} (\bibinfo {year} {2012})}\BibitemShut
  {NoStop}%
\bibitem [{\citenamefont {Guti{\'{e}}rrez}\ \emph {et~al.}(2013)\citenamefont
  {Guti{\'{e}}rrez}, \citenamefont {Perea-L{\'{o}}pez}, \citenamefont
  {El{\'{i}}as}, \citenamefont {Berkdemir}, \citenamefont {Wang}, \citenamefont
  {Lv}, \citenamefont {L{\'{o}}pez-Ur{\'{i}}as}, \citenamefont {Crespi},
  \citenamefont {Terrones},\ and\ \citenamefont {Terrones}}]{Gut2013}%
  \BibitemOpen
  \bibfield  {author} {\bibinfo {author} {\bibfnamefont {H.~R.}\ \bibnamefont
  {Guti{\'{e}}rrez}}, \bibinfo {author} {\bibfnamefont {N.}~\bibnamefont
  {Perea-L{\'{o}}pez}}, \bibinfo {author} {\bibfnamefont {A.~L.}\ \bibnamefont
  {El{\'{i}}as}}, \bibinfo {author} {\bibfnamefont {A.}~\bibnamefont
  {Berkdemir}}, \bibinfo {author} {\bibfnamefont {B.}~\bibnamefont {Wang}},
  \bibinfo {author} {\bibfnamefont {R.}~\bibnamefont {Lv}}, \bibinfo {author}
  {\bibfnamefont {F.}~\bibnamefont {L{\'{o}}pez-Ur{\'{i}}as}}, \bibinfo
  {author} {\bibfnamefont {V.~H.}\ \bibnamefont {Crespi}}, \bibinfo {author}
  {\bibfnamefont {H.}~\bibnamefont {Terrones}}, \ and\ \bibinfo {author}
  {\bibfnamefont {M.}~\bibnamefont {Terrones}},\ }\href {\doibase
  10.1021/nl3026357} {\bibfield  {journal} {\bibinfo  {journal} {Nano Lett.}\
  }\textbf {\bibinfo {volume} {13}},\ \bibinfo {pages} {3447} (\bibinfo {year}
  {2013})}\BibitemShut {NoStop}%
\bibitem [{\citenamefont {Zhao}\ \emph {et~al.}(2013)\citenamefont {Zhao},
  \citenamefont {Ghorannevis}, \citenamefont {Chu}, \citenamefont {Toh},
  \citenamefont {Kloc}, \citenamefont {Tan},\ and\ \citenamefont
  {Eda}}]{Zhao2013}%
  \BibitemOpen
  \bibfield  {author} {\bibinfo {author} {\bibfnamefont {W.}~\bibnamefont
  {Zhao}}, \bibinfo {author} {\bibfnamefont {Z.}~\bibnamefont {Ghorannevis}},
  \bibinfo {author} {\bibfnamefont {L.}~\bibnamefont {Chu}}, \bibinfo {author}
  {\bibfnamefont {M.}~\bibnamefont {Toh}}, \bibinfo {author} {\bibfnamefont
  {C.}~\bibnamefont {Kloc}}, \bibinfo {author} {\bibfnamefont {P.-H.}\
  \bibnamefont {Tan}}, \ and\ \bibinfo {author} {\bibfnamefont
  {G.}~\bibnamefont {Eda}},\ }\href {\doibase 10.1021/nn305275h} {\bibfield
  {journal} {\bibinfo  {journal} {ACS Nano}\ }\textbf {\bibinfo {volume} {7}},\
  \bibinfo {pages} {791} (\bibinfo {year} {2013})}\BibitemShut {NoStop}%
\bibitem [{\citenamefont {Wang}\ \emph {et~al.}(2012)\citenamefont {Wang},
  \citenamefont {Kalantar-Zadeh}, \citenamefont {Kis}, \citenamefont
  {Coleman},\ and\ \citenamefont {Strano}}]{Wang2012}%
  \BibitemOpen
  \bibfield  {author} {\bibinfo {author} {\bibfnamefont {Q.~H.}\ \bibnamefont
  {Wang}}, \bibinfo {author} {\bibfnamefont {K.}~\bibnamefont
  {Kalantar-Zadeh}}, \bibinfo {author} {\bibfnamefont {A.}~\bibnamefont {Kis}},
  \bibinfo {author} {\bibfnamefont {J.~N.}\ \bibnamefont {Coleman}}, \ and\
  \bibinfo {author} {\bibfnamefont {M.~S.}\ \bibnamefont {Strano}},\ }\href
  {\doibase 10.1038/nnano.2012.193} {\bibfield  {journal} {\bibinfo  {journal}
  {Nat. Nanotechnol.}\ }\textbf {\bibinfo {volume} {7}},\ \bibinfo {pages}
  {699} (\bibinfo {year} {2012})}\BibitemShut {NoStop}%
\bibitem [{\citenamefont {Chhowalla}\ \emph {et~al.}(2013)\citenamefont
  {Chhowalla}, \citenamefont {Shin}, \citenamefont {Eda}, \citenamefont {Li},
  \citenamefont {Loh},\ and\ \citenamefont {Zhang}}]{Chhowalla2013}%
  \BibitemOpen
  \bibfield  {author} {\bibinfo {author} {\bibfnamefont {M.}~\bibnamefont
  {Chhowalla}}, \bibinfo {author} {\bibfnamefont {H.~S.}\ \bibnamefont {Shin}},
  \bibinfo {author} {\bibfnamefont {G.}~\bibnamefont {Eda}}, \bibinfo {author}
  {\bibfnamefont {L.-J.}\ \bibnamefont {Li}}, \bibinfo {author} {\bibfnamefont
  {K.~P.}\ \bibnamefont {Loh}}, \ and\ \bibinfo {author} {\bibfnamefont
  {H.}~\bibnamefont {Zhang}},\ }\href {\doibase 10.1038/nchem.1589} {\bibfield
  {journal} {\bibinfo  {journal} {Nat. Chem.}\ }\textbf {\bibinfo {volume}
  {5}},\ \bibinfo {pages} {263} (\bibinfo {year} {2013})}\BibitemShut {NoStop}%
\bibitem [{\citenamefont {Zeng}\ \emph {et~al.}(2012)\citenamefont {Zeng},
  \citenamefont {Dai}, \citenamefont {Yao}, \citenamefont {Xiao},\ and\
  \citenamefont {Cui}}]{Zeng}%
  \BibitemOpen
  \bibfield  {author} {\bibinfo {author} {\bibfnamefont {H.}~\bibnamefont
  {Zeng}}, \bibinfo {author} {\bibfnamefont {J.}~\bibnamefont {Dai}}, \bibinfo
  {author} {\bibfnamefont {W.}~\bibnamefont {Yao}}, \bibinfo {author}
  {\bibfnamefont {D.}~\bibnamefont {Xiao}}, \ and\ \bibinfo {author}
  {\bibfnamefont {X.}~\bibnamefont {Cui}},\ }\href {\doibase
  10.1038/nnano.2012.95} {\bibfield  {journal} {\bibinfo  {journal} {Nat.
  Nanotechnol.}\ }\textbf {\bibinfo {volume} {7}},\ \bibinfo {pages} {490}
  (\bibinfo {year} {2012})}\BibitemShut {NoStop}%
\bibitem [{\citenamefont {Mak}\ \emph {et~al.}(2012)\citenamefont {Mak},
  \citenamefont {He}, \citenamefont {Shan},\ and\ \citenamefont {Heinz}}]{Mak}%
  \BibitemOpen
  \bibfield  {author} {\bibinfo {author} {\bibfnamefont {K.~F.}\ \bibnamefont
  {Mak}}, \bibinfo {author} {\bibfnamefont {K.}~\bibnamefont {He}}, \bibinfo
  {author} {\bibfnamefont {J.}~\bibnamefont {Shan}}, \ and\ \bibinfo {author}
  {\bibfnamefont {T.~F.}\ \bibnamefont {Heinz}},\ }\href {\doibase
  10.1038/nnano.2012.96} {\bibfield  {journal} {\bibinfo  {journal} {Nat.
  Nanotechnol.}\ }\textbf {\bibinfo {volume} {7}},\ \bibinfo {pages} {494}
  (\bibinfo {year} {2012})}\BibitemShut {NoStop}%
\bibitem [{\citenamefont {Cao}\ \emph {et~al.}(2012)\citenamefont {Cao},
  \citenamefont {Wang}, \citenamefont {Han}, \citenamefont {Ye}, \citenamefont
  {Zhu}, \citenamefont {Shi}, \citenamefont {Niu}, \citenamefont {Tan},
  \citenamefont {Wang}, \citenamefont {Liu},\ and\ \citenamefont {Feng}}]{Cao}%
  \BibitemOpen
  \bibfield  {author} {\bibinfo {author} {\bibfnamefont {T.}~\bibnamefont
  {Cao}}, \bibinfo {author} {\bibfnamefont {G.}~\bibnamefont {Wang}}, \bibinfo
  {author} {\bibfnamefont {W.}~\bibnamefont {Han}}, \bibinfo {author}
  {\bibfnamefont {H.}~\bibnamefont {Ye}}, \bibinfo {author} {\bibfnamefont
  {C.}~\bibnamefont {Zhu}}, \bibinfo {author} {\bibfnamefont {J.}~\bibnamefont
  {Shi}}, \bibinfo {author} {\bibfnamefont {Q.}~\bibnamefont {Niu}}, \bibinfo
  {author} {\bibfnamefont {P.}~\bibnamefont {Tan}}, \bibinfo {author}
  {\bibfnamefont {E.}~\bibnamefont {Wang}}, \bibinfo {author} {\bibfnamefont
  {B.}~\bibnamefont {Liu}}, \ and\ \bibinfo {author} {\bibfnamefont
  {J.}~\bibnamefont {Feng}},\ }\href {\doibase 10.1038/ncomms1882} {\bibfield
  {journal} {\bibinfo  {journal} {Nat. Commun.}\ }\textbf {\bibinfo {volume}
  {3}},\ \bibinfo {pages} {887} (\bibinfo {year} {2012})}\BibitemShut {NoStop}%
\bibitem [{\citenamefont {Yu}\ \emph {et~al.}(2015)\citenamefont {Yu},
  \citenamefont {Cui}, \citenamefont {Xu},\ and\ \citenamefont {Yao}}]{Yu}%
  \BibitemOpen
  \bibfield  {author} {\bibinfo {author} {\bibfnamefont {H.}~\bibnamefont
  {Yu}}, \bibinfo {author} {\bibfnamefont {X.}~\bibnamefont {Cui}}, \bibinfo
  {author} {\bibfnamefont {X.}~\bibnamefont {Xu}}, \ and\ \bibinfo {author}
  {\bibfnamefont {W.}~\bibnamefont {Yao}},\ }\href {\doibase
  10.1093/nsr/nwu078} {\bibfield  {journal} {\bibinfo  {journal} {Natl. Sci.
  Rev.}\ }\textbf {\bibinfo {volume} {2}},\ \bibinfo {pages} {57} (\bibinfo
  {year} {2015})}\BibitemShut {NoStop}%
\bibitem [{\citenamefont {Sinova}\ \emph {et~al.}(2015)\citenamefont {Sinova},
  \citenamefont {Valenzuela}, \citenamefont {Wunderlich}, \citenamefont
  {Back},\ and\ \citenamefont {Jungwirth}}]{Sinova}%
  \BibitemOpen
  \bibfield  {author} {\bibinfo {author} {\bibfnamefont {J.}~\bibnamefont
  {Sinova}}, \bibinfo {author} {\bibfnamefont {S.~O.}\ \bibnamefont
  {Valenzuela}}, \bibinfo {author} {\bibfnamefont {J.}~\bibnamefont
  {Wunderlich}}, \bibinfo {author} {\bibfnamefont {C.~H.}\ \bibnamefont
  {Back}}, \ and\ \bibinfo {author} {\bibfnamefont {T.}~\bibnamefont
  {Jungwirth}},\ }\href {\doibase 10.1103/revmodphys.87.1213} {\bibfield
  {journal} {\bibinfo  {journal} {Rev. Mod. Phys.}\ }\textbf {\bibinfo {volume}
  {87}},\ \bibinfo {pages} {1213} (\bibinfo {year} {2015})}\BibitemShut
  {NoStop}%
\bibitem [{\citenamefont {Hai}(2020)}]{Hai}%
  \BibitemOpen
  \bibfield  {author} {\bibinfo {author} {\bibfnamefont {P.~N.}\ \bibnamefont
  {Hai}},\ }\href {\doibase 10.3379/msjmag.2009rv001} {\bibfield  {journal}
  {\bibinfo  {journal} {J. Magn. Soc. Jpn.}\ }\textbf {\bibinfo {volume}
  {44}},\ \bibinfo {pages} {137} (\bibinfo {year} {2020})}\BibitemShut
  {NoStop}%
\bibitem [{\citenamefont {Kato}\ \emph {et~al.}(2004)\citenamefont {Kato},
  \citenamefont {Myers}, \citenamefont {Gossard},\ and\ \citenamefont
  {Awschalom}}]{Kato}%
  \BibitemOpen
  \bibfield  {author} {\bibinfo {author} {\bibfnamefont {Y.~K.}\ \bibnamefont
  {Kato}}, \bibinfo {author} {\bibfnamefont {R.~C.}\ \bibnamefont {Myers}},
  \bibinfo {author} {\bibfnamefont {A.~C.}\ \bibnamefont {Gossard}}, \ and\
  \bibinfo {author} {\bibfnamefont {D.~D.}\ \bibnamefont {Awschalom}},\ }\href
  {\doibase 10.1126/science.1105514} {\bibfield  {journal} {\bibinfo  {journal}
  {Science}\ }\textbf {\bibinfo {volume} {306}},\ \bibinfo {pages} {1910}
  (\bibinfo {year} {2004})}\BibitemShut {NoStop}%
\bibitem [{\citenamefont {Wunderlich}\ \emph {et~al.}(2005)\citenamefont
  {Wunderlich}, \citenamefont {Kaestner}, \citenamefont {Sinova},\ and\
  \citenamefont {Jungwirth}}]{Wun}%
  \BibitemOpen
  \bibfield  {author} {\bibinfo {author} {\bibfnamefont {J.}~\bibnamefont
  {Wunderlich}}, \bibinfo {author} {\bibfnamefont {B.}~\bibnamefont
  {Kaestner}}, \bibinfo {author} {\bibfnamefont {J.}~\bibnamefont {Sinova}}, \
  and\ \bibinfo {author} {\bibfnamefont {T.}~\bibnamefont {Jungwirth}},\ }\href
  {\doibase 10.1103/physrevlett.94.047204} {\bibfield  {journal} {\bibinfo
  {journal} {Phys. Rev. Lett.}\ }\textbf {\bibinfo {volume} {94}},\ \bibinfo
  {pages} {047204} (\bibinfo {year} {2005})}\BibitemShut {NoStop}%
\bibitem [{\citenamefont {Kim}\ and\ \citenamefont {Son}(2017)}]{Kim}%
  \BibitemOpen
  \bibfield  {author} {\bibinfo {author} {\bibfnamefont {S.}~\bibnamefont
  {Kim}}\ and\ \bibinfo {author} {\bibfnamefont {Y.-W.}\ \bibnamefont {Son}},\
  }\href {\doibase 10.1103/physrevb.96.155439} {\bibfield  {journal} {\bibinfo
  {journal} {Phys. Rev. B}\ }\textbf {\bibinfo {volume} {96}},\ \bibinfo
  {pages} {155439} (\bibinfo {year} {2017})}\BibitemShut {NoStop}%
\bibitem [{\citenamefont {He}\ \emph {et~al.}(2018)\citenamefont {He},
  \citenamefont {Zhou}, \citenamefont {He}, \citenamefont {Yuan}, \citenamefont
  {Zhang},\ and\ \citenamefont {Law}}]{He}%
  \BibitemOpen
  \bibfield  {author} {\bibinfo {author} {\bibfnamefont {W.-Y.}\ \bibnamefont
  {He}}, \bibinfo {author} {\bibfnamefont {B.~T.}\ \bibnamefont {Zhou}},
  \bibinfo {author} {\bibfnamefont {J.~J.}\ \bibnamefont {He}}, \bibinfo
  {author} {\bibfnamefont {N.~F.~Q.}\ \bibnamefont {Yuan}}, \bibinfo {author}
  {\bibfnamefont {T.}~\bibnamefont {Zhang}}, \ and\ \bibinfo {author}
  {\bibfnamefont {K.~T.}\ \bibnamefont {Law}},\ }\href {\doibase
  10.1038/s42005-018-0041-4} {\bibfield  {journal} {\bibinfo  {journal}
  {Commun. Phys.}\ }\textbf {\bibinfo {volume} {1}},\ \bibinfo {pages} {40}
  (\bibinfo {year} {2018})}\BibitemShut {NoStop}%
\bibitem [{\citenamefont {Xi}\ \emph {et~al.}(2016)\citenamefont {Xi},
  \citenamefont {Wang}, \citenamefont {Zhao}, \citenamefont {Park},
  \citenamefont {Law}, \citenamefont {Berger}, \citenamefont {Forró},
  \citenamefont {Shan},\ and\ \citenamefont {Mak}}]{Xi}%
  \BibitemOpen
  \bibfield  {author} {\bibinfo {author} {\bibfnamefont {X.}~\bibnamefont
  {Xi}}, \bibinfo {author} {\bibfnamefont {Z.}~\bibnamefont {Wang}}, \bibinfo
  {author} {\bibfnamefont {W.}~\bibnamefont {Zhao}}, \bibinfo {author}
  {\bibfnamefont {J.-H.}\ \bibnamefont {Park}}, \bibinfo {author}
  {\bibfnamefont {K.~T.}\ \bibnamefont {Law}}, \bibinfo {author} {\bibfnamefont
  {H.}~\bibnamefont {Berger}}, \bibinfo {author} {\bibfnamefont
  {L.}~\bibnamefont {Forró}}, \bibinfo {author} {\bibfnamefont
  {J.}~\bibnamefont {Shan}}, \ and\ \bibinfo {author} {\bibfnamefont {K.~F.}\
  \bibnamefont {Mak}},\ }\href {\doibase 10.1038/nphys3538} {\bibfield
  {journal} {\bibinfo  {journal} {Nat. Phys.}\ }\textbf {\bibinfo {volume}
  {12}},\ \bibinfo {pages} {139} (\bibinfo {year} {2016})}\BibitemShut
  {NoStop}%
\bibitem [{\citenamefont {Sohn}\ \emph {et~al.}(2018)\citenamefont {Sohn},
  \citenamefont {Xi}, \citenamefont {He}, \citenamefont {Jiang}, \citenamefont
  {Wang}, \citenamefont {Kang}, \citenamefont {Park}, \citenamefont {Berger},
  \citenamefont {Forró}, \citenamefont {Law}, \citenamefont {Shan},\ and\
  \citenamefont {Mak}}]{Sohn}%
  \BibitemOpen
  \bibfield  {author} {\bibinfo {author} {\bibfnamefont {E.}~\bibnamefont
  {Sohn}}, \bibinfo {author} {\bibfnamefont {X.}~\bibnamefont {Xi}}, \bibinfo
  {author} {\bibfnamefont {W.-Y.}\ \bibnamefont {He}}, \bibinfo {author}
  {\bibfnamefont {S.}~\bibnamefont {Jiang}}, \bibinfo {author} {\bibfnamefont
  {Z.}~\bibnamefont {Wang}}, \bibinfo {author} {\bibfnamefont {K.}~\bibnamefont
  {Kang}}, \bibinfo {author} {\bibfnamefont {J.-H.}\ \bibnamefont {Park}},
  \bibinfo {author} {\bibfnamefont {H.}~\bibnamefont {Berger}}, \bibinfo
  {author} {\bibfnamefont {L.}~\bibnamefont {Forró}}, \bibinfo {author}
  {\bibfnamefont {K.~T.}\ \bibnamefont {Law}}, \bibinfo {author} {\bibfnamefont
  {J.}~\bibnamefont {Shan}}, \ and\ \bibinfo {author} {\bibfnamefont {K.~F.}\
  \bibnamefont {Mak}},\ }\href {\doibase 10.1038/s41563-018-0061-1} {\bibfield
  {journal} {\bibinfo  {journal} {Nat. Mater.}\ }\textbf {\bibinfo {volume}
  {17}},\ \bibinfo {pages} {504} (\bibinfo {year} {2018})}\BibitemShut
  {NoStop}%
\bibitem [{\citenamefont {Lu}\ \emph {et~al.}(2015)\citenamefont {Lu},
  \citenamefont {Zheliuk}, \citenamefont {Leermakers}, \citenamefont {Yuan},
  \citenamefont {Zeitler}, \citenamefont {Law},\ and\ \citenamefont {Ye}}]{Lu}%
  \BibitemOpen
  \bibfield  {author} {\bibinfo {author} {\bibfnamefont {J.~M.}\ \bibnamefont
  {Lu}}, \bibinfo {author} {\bibfnamefont {O.}~\bibnamefont {Zheliuk}},
  \bibinfo {author} {\bibfnamefont {I.}~\bibnamefont {Leermakers}}, \bibinfo
  {author} {\bibfnamefont {N.~F.~Q.}\ \bibnamefont {Yuan}}, \bibinfo {author}
  {\bibfnamefont {U.}~\bibnamefont {Zeitler}}, \bibinfo {author} {\bibfnamefont
  {K.~T.}\ \bibnamefont {Law}}, \ and\ \bibinfo {author} {\bibfnamefont
  {J.~T.}\ \bibnamefont {Ye}},\ }\href {\doibase 10.1126/science.aab2277}
  {\bibfield  {journal} {\bibinfo  {journal} {Science}\ }\textbf {\bibinfo
  {volume} {350}},\ \bibinfo {pages} {1353} (\bibinfo {year}
  {2015})}\BibitemShut {NoStop}%
\bibitem [{\citenamefont {Saito}\ \emph {et~al.}(2016)\citenamefont {Saito},
  \citenamefont {Nakamura}, \citenamefont {Bahramy}, \citenamefont {Kohama},
  \citenamefont {Ye}, \citenamefont {Kasahara}, \citenamefont {Nakagawa},
  \citenamefont {Onga}, \citenamefont {Tokunaga}, \citenamefont {Nojima},
  \citenamefont {Yanase},\ and\ \citenamefont {Iwasa}}]{Saito}%
  \BibitemOpen
  \bibfield  {author} {\bibinfo {author} {\bibfnamefont {Y.}~\bibnamefont
  {Saito}}, \bibinfo {author} {\bibfnamefont {Y.}~\bibnamefont {Nakamura}},
  \bibinfo {author} {\bibfnamefont {M.~S.}\ \bibnamefont {Bahramy}}, \bibinfo
  {author} {\bibfnamefont {Y.}~\bibnamefont {Kohama}}, \bibinfo {author}
  {\bibfnamefont {J.}~\bibnamefont {Ye}}, \bibinfo {author} {\bibfnamefont
  {Y.}~\bibnamefont {Kasahara}}, \bibinfo {author} {\bibfnamefont
  {Y.}~\bibnamefont {Nakagawa}}, \bibinfo {author} {\bibfnamefont
  {M.}~\bibnamefont {Onga}}, \bibinfo {author} {\bibfnamefont {M.}~\bibnamefont
  {Tokunaga}}, \bibinfo {author} {\bibfnamefont {T.}~\bibnamefont {Nojima}},
  \bibinfo {author} {\bibfnamefont {Y.}~\bibnamefont {Yanase}}, \ and\ \bibinfo
  {author} {\bibfnamefont {Y.}~\bibnamefont {Iwasa}},\ }\href {\doibase
  10.1038/nphys3580} {\bibfield  {journal} {\bibinfo  {journal} {Nat. Phys.}\
  }\textbf {\bibinfo {volume} {12}},\ \bibinfo {pages} {144} (\bibinfo {year}
  {2016})}\BibitemShut {NoStop}%
\bibitem [{\citenamefont {Zhou}\ \emph {et~al.}(2016)\citenamefont {Zhou},
  \citenamefont {Yuan}, \citenamefont {Jiang},\ and\ \citenamefont
  {Law}}]{Zhou}%
  \BibitemOpen
  \bibfield  {author} {\bibinfo {author} {\bibfnamefont {B.~T.}\ \bibnamefont
  {Zhou}}, \bibinfo {author} {\bibfnamefont {N.~F.~Q.}\ \bibnamefont {Yuan}},
  \bibinfo {author} {\bibfnamefont {H.-L.}\ \bibnamefont {Jiang}}, \ and\
  \bibinfo {author} {\bibfnamefont {K.~T.}\ \bibnamefont {Law}},\ }\href
  {\doibase 10.1103/physrevb.93.180501} {\bibfield  {journal} {\bibinfo
  {journal} {Phys. Rev. B}\ }\textbf {\bibinfo {volume} {93}},\ \bibinfo
  {pages} {180501(R)} (\bibinfo {year} {2016})}\BibitemShut {NoStop}%
\bibitem [{\citenamefont {Bawden}\ \emph {et~al.}(2016)\citenamefont {Bawden},
  \citenamefont {Cooil}, \citenamefont {Mazzola}, \citenamefont {Riley},
  \citenamefont {Collins-McIntyre}, \citenamefont {Sunko}, \citenamefont
  {Hunvik}, \citenamefont {Leandersson}, \citenamefont {Polley}, \citenamefont
  {Balasubramanian}, \citenamefont {Kim}, \citenamefont {Hoesch}, \citenamefont
  {Wells}, \citenamefont {Balakrishnan}, \citenamefont {Bahramy},\ and\
  \citenamefont {King}}]{Baw}%
  \BibitemOpen
  \bibfield  {author} {\bibinfo {author} {\bibfnamefont {L.}~\bibnamefont
  {Bawden}}, \bibinfo {author} {\bibfnamefont {S.~P.}\ \bibnamefont {Cooil}},
  \bibinfo {author} {\bibfnamefont {F.}~\bibnamefont {Mazzola}}, \bibinfo
  {author} {\bibfnamefont {J.~M.}\ \bibnamefont {Riley}}, \bibinfo {author}
  {\bibfnamefont {L.~J.}\ \bibnamefont {Collins-McIntyre}}, \bibinfo {author}
  {\bibfnamefont {V.}~\bibnamefont {Sunko}}, \bibinfo {author} {\bibfnamefont
  {K.~W.~B.}\ \bibnamefont {Hunvik}}, \bibinfo {author} {\bibfnamefont
  {M.}~\bibnamefont {Leandersson}}, \bibinfo {author} {\bibfnamefont {C.~M.}\
  \bibnamefont {Polley}}, \bibinfo {author} {\bibfnamefont {T.}~\bibnamefont
  {Balasubramanian}}, \bibinfo {author} {\bibfnamefont {T.~K.}\ \bibnamefont
  {Kim}}, \bibinfo {author} {\bibfnamefont {M.}~\bibnamefont {Hoesch}},
  \bibinfo {author} {\bibfnamefont {J.~W.}\ \bibnamefont {Wells}}, \bibinfo
  {author} {\bibfnamefont {G.}~\bibnamefont {Balakrishnan}}, \bibinfo {author}
  {\bibfnamefont {M.~S.}\ \bibnamefont {Bahramy}}, \ and\ \bibinfo {author}
  {\bibfnamefont {P.~D.~C.}\ \bibnamefont {King}},\ }\href {\doibase
  10.1038/ncomms11711} {\bibfield  {journal} {\bibinfo  {journal} {Nat.
  Commun.}\ }\textbf {\bibinfo {volume} {7}},\ \bibinfo {pages} {11711}
  (\bibinfo {year} {2016})}\BibitemShut {NoStop}%
\bibitem [{\citenamefont {Liu}\ \emph {et~al.}(2013)\citenamefont {Liu},
  \citenamefont {Shan}, \citenamefont {Yao}, \citenamefont {Yao},\ and\
  \citenamefont {Xiao}}]{Liu}%
  \BibitemOpen
  \bibfield  {author} {\bibinfo {author} {\bibfnamefont {G.-B.}\ \bibnamefont
  {Liu}}, \bibinfo {author} {\bibfnamefont {W.-Y.}\ \bibnamefont {Shan}},
  \bibinfo {author} {\bibfnamefont {Y.}~\bibnamefont {Yao}}, \bibinfo {author}
  {\bibfnamefont {W.}~\bibnamefont {Yao}}, \ and\ \bibinfo {author}
  {\bibfnamefont {D.}~\bibnamefont {Xiao}},\ }\href {\doibase
  10.1103/physrevb.88.085433} {\bibfield  {journal} {\bibinfo  {journal} {Phys.
  Rev. B}\ }\textbf {\bibinfo {volume} {88}},\ \bibinfo {pages} {085433}
  (\bibinfo {year} {2013})}\BibitemShut {NoStop}%
\bibitem [{sup()}]{supplementary}%
  \BibitemOpen
  \href@noop {} {\ }\bibinfo {note} {See Supplemetary Material at [URL will be
  inserted by publisher] for details of matrix elements in an effective
  Hamiltonian. It is also presented the doping and temperature effects on
  optical spin Hall conductivity.}\BibitemShut {Stop}%
\bibitem [{\citenamefont {Qiao}\ \emph {et~al.}(2018)\citenamefont {Qiao},
  \citenamefont {Zhou}, \citenamefont {Yuan},\ and\ \citenamefont
  {Zhao}}]{Qiao}%
  \BibitemOpen
  \bibfield  {author} {\bibinfo {author} {\bibfnamefont {J.}~\bibnamefont
  {Qiao}}, \bibinfo {author} {\bibfnamefont {J.}~\bibnamefont {Zhou}}, \bibinfo
  {author} {\bibfnamefont {Z.}~\bibnamefont {Yuan}}, \ and\ \bibinfo {author}
  {\bibfnamefont {W.}~\bibnamefont {Zhao}},\ }\href {\doibase
  10.1103/physrevb.98.214402} {\bibfield  {journal} {\bibinfo  {journal} {Phys.
  Rev. B}\ }\textbf {\bibinfo {volume} {98}},\ \bibinfo {pages} {214402}
  (\bibinfo {year} {2018})}\BibitemShut {NoStop}%
\bibitem [{\citenamefont {Guo}\ \emph {et~al.}(2005)\citenamefont {Guo},
  \citenamefont {Yao},\ and\ \citenamefont {Niu}}]{Guo}%
  \BibitemOpen
  \bibfield  {author} {\bibinfo {author} {\bibfnamefont {G.~Y.}\ \bibnamefont
  {Guo}}, \bibinfo {author} {\bibfnamefont {Y.}~\bibnamefont {Yao}}, \ and\
  \bibinfo {author} {\bibfnamefont {Q.}~\bibnamefont {Niu}},\ }\href {\doibase
  10.1103/physrevlett.94.226601} {\bibfield  {journal} {\bibinfo  {journal}
  {Phys. Rev. Lett.}\ }\textbf {\bibinfo {volume} {94}},\ \bibinfo {pages}
  {226601} (\bibinfo {year} {2005})}\BibitemShut {NoStop}%
\bibitem [{\citenamefont {Sengupta}\ \emph {et~al.}(2015)\citenamefont
  {Sengupta}, \citenamefont {Rakheja},\ and\ \citenamefont {Bellotti}}]{Seng}%
  \BibitemOpen
  \bibfield  {author} {\bibinfo {author} {\bibfnamefont {P.}~\bibnamefont
  {Sengupta}}, \bibinfo {author} {\bibfnamefont {S.}~\bibnamefont {Rakheja}}, \
  and\ \bibinfo {author} {\bibfnamefont {E.}~\bibnamefont {Bellotti}},\
  }\href@noop {} {\bibfield  {journal} {\bibinfo  {journal} {arXiv:1512.06734}\
  } (\bibinfo {year} {2015})}\BibitemShut {NoStop}%
\bibitem [{\citenamefont {Vargiamidis}\ \emph {et~al.}(2014)\citenamefont
  {Vargiamidis}, \citenamefont {Vasilopoulos},\ and\ \citenamefont
  {Hai}}]{Varg}%
  \BibitemOpen
  \bibfield  {author} {\bibinfo {author} {\bibfnamefont {V.}~\bibnamefont
  {Vargiamidis}}, \bibinfo {author} {\bibfnamefont {P.}~\bibnamefont
  {Vasilopoulos}}, \ and\ \bibinfo {author} {\bibfnamefont {G.-Q.}\
  \bibnamefont {Hai}},\ }\href {\doibase 10.1088/0953-8984/26/34/345303}
  {\bibfield  {journal} {\bibinfo  {journal} {J. Phys. Condens. Matter}\
  }\textbf {\bibinfo {volume} {26}},\ \bibinfo {pages} {345303} (\bibinfo
  {year} {2014})}\BibitemShut {NoStop}%
\bibitem [{\citenamefont {Tanaka}\ \emph {et~al.}(2008)\citenamefont {Tanaka},
  \citenamefont {Kontani}, \citenamefont {Naito}, \citenamefont {Naito},
  \citenamefont {Hirashima}, \citenamefont {Yamada},\ and\ \citenamefont
  {Inoue}}]{Tanaka}%
  \BibitemOpen
  \bibfield  {author} {\bibinfo {author} {\bibfnamefont {T.}~\bibnamefont
  {Tanaka}}, \bibinfo {author} {\bibfnamefont {H.}~\bibnamefont {Kontani}},
  \bibinfo {author} {\bibfnamefont {M.}~\bibnamefont {Naito}}, \bibinfo
  {author} {\bibfnamefont {T.}~\bibnamefont {Naito}}, \bibinfo {author}
  {\bibfnamefont {D.~S.}\ \bibnamefont {Hirashima}}, \bibinfo {author}
  {\bibfnamefont {K.}~\bibnamefont {Yamada}}, \ and\ \bibinfo {author}
  {\bibfnamefont {J.}~\bibnamefont {Inoue}},\ }\href {\doibase
  10.1103/physrevb.77.165117} {\bibfield  {journal} {\bibinfo  {journal} {Phys.
  Rev. B}\ }\textbf {\bibinfo {volume} {77}},\ \bibinfo {pages} {165117}
  (\bibinfo {year} {2008})}\BibitemShut {NoStop}%
\bibitem [{\citenamefont {Ferreira}\ \emph {et~al.}(2011)\citenamefont
  {Ferreira}, \citenamefont {Viana-Gomes}, \citenamefont {Bludov},
  \citenamefont {Pereira}, \citenamefont {Peres},\ and\ \citenamefont
  {Castro~Neto}}]{Ferreira}%
  \BibitemOpen
  \bibfield  {author} {\bibinfo {author} {\bibfnamefont {A.}~\bibnamefont
  {Ferreira}}, \bibinfo {author} {\bibfnamefont {J.}~\bibnamefont
  {Viana-Gomes}}, \bibinfo {author} {\bibfnamefont {Y.~V.}\ \bibnamefont
  {Bludov}}, \bibinfo {author} {\bibfnamefont {V.}~\bibnamefont {Pereira}},
  \bibinfo {author} {\bibfnamefont {N.~M.~R.}\ \bibnamefont {Peres}}, \ and\
  \bibinfo {author} {\bibfnamefont {A.~H.}\ \bibnamefont {Castro~Neto}},\
  }\href {\doibase 10.1103/physrevb.84.235410} {\bibfield  {journal} {\bibinfo
  {journal} {Phys. Rev. B}\ }\textbf {\bibinfo {volume} {84}},\ \bibinfo
  {pages} {235410} (\bibinfo {year} {2011})}\BibitemShut {NoStop}%
\bibitem [{\citenamefont {Morimoto}\ \emph {et~al.}(2009)\citenamefont
  {Morimoto}, \citenamefont {Hatsugai},\ and\ \citenamefont {Aoki}}]{Morimoto}%
  \BibitemOpen
  \bibfield  {author} {\bibinfo {author} {\bibfnamefont {T.}~\bibnamefont
  {Morimoto}}, \bibinfo {author} {\bibfnamefont {Y.}~\bibnamefont {Hatsugai}},
  \ and\ \bibinfo {author} {\bibfnamefont {H.}~\bibnamefont {Aoki}},\ }\href
  {\doibase 10.1103/physrevlett.103.116803} {\bibfield  {journal} {\bibinfo
  {journal} {Phys. Rev. Lett.}\ }\textbf {\bibinfo {volume} {103}},\ \bibinfo
  {pages} {116803} (\bibinfo {year} {2009})}\BibitemShut {NoStop}%
\bibitem [{\citenamefont {Yao}\ \emph {et~al.}(2004)\citenamefont {Yao},
  \citenamefont {Kleinman}, \citenamefont {MacDonald}, \citenamefont {Sinova},
  \citenamefont {Jungwirth}, \citenamefont {Wang}, \citenamefont {Wang},\ and\
  \citenamefont {Niu}}]{Yugui}%
  \BibitemOpen
  \bibfield  {author} {\bibinfo {author} {\bibfnamefont {Y.}~\bibnamefont
  {Yao}}, \bibinfo {author} {\bibfnamefont {L.}~\bibnamefont {Kleinman}},
  \bibinfo {author} {\bibfnamefont {A.~H.}\ \bibnamefont {MacDonald}}, \bibinfo
  {author} {\bibfnamefont {J.}~\bibnamefont {Sinova}}, \bibinfo {author}
  {\bibfnamefont {T.}~\bibnamefont {Jungwirth}}, \bibinfo {author}
  {\bibfnamefont {D.-s.}\ \bibnamefont {Wang}}, \bibinfo {author}
  {\bibfnamefont {E.}~\bibnamefont {Wang}}, \ and\ \bibinfo {author}
  {\bibfnamefont {Q.}~\bibnamefont {Niu}},\ }\href {\doibase
  10.1103/physrevlett.92.037204} {\bibfield  {journal} {\bibinfo  {journal}
  {Phys. Rev. Lett.}\ }\textbf {\bibinfo {volume} {92}},\ \bibinfo {pages}
  {037204} (\bibinfo {year} {2004})}\BibitemShut {NoStop}%
\bibitem [{\citenamefont {Li}\ and\ \citenamefont {Carbotte}(2012)}]{LiZ}%
  \BibitemOpen
  \bibfield  {author} {\bibinfo {author} {\bibfnamefont {Z.}~\bibnamefont
  {Li}}\ and\ \bibinfo {author} {\bibfnamefont {J.~P.}\ \bibnamefont
  {Carbotte}},\ }\href {\doibase 10.1103/physrevb.86.205425} {\bibfield
  {journal} {\bibinfo  {journal} {Phys. Rev. B}\ }\textbf {\bibinfo {volume}
  {86}},\ \bibinfo {pages} {205425} (\bibinfo {year} {2012})}\BibitemShut
  {NoStop}%
\bibitem [{\citenamefont {Akita}\ \emph {et~al.}(2020)\citenamefont {Akita},
  \citenamefont {Fujii}, \citenamefont {Maruyama}, \citenamefont {Okada},\ and\
  \citenamefont {Wakabayashi}}]{Akita}%
  \BibitemOpen
  \bibfield  {author} {\bibinfo {author} {\bibfnamefont {M.}~\bibnamefont
  {Akita}}, \bibinfo {author} {\bibfnamefont {Y.}~\bibnamefont {Fujii}},
  \bibinfo {author} {\bibfnamefont {M.}~\bibnamefont {Maruyama}}, \bibinfo
  {author} {\bibfnamefont {S.}~\bibnamefont {Okada}}, \ and\ \bibinfo {author}
  {\bibfnamefont {K.}~\bibnamefont {Wakabayashi}},\ }\href {\doibase
  10.1103/physrevb.101.085418} {\bibfield  {journal} {\bibinfo  {journal}
  {Phys. Rev. B}\ }\textbf {\bibinfo {volume} {101}},\ \bibinfo {pages}
  {085418} (\bibinfo {year} {2020})}\BibitemShut {NoStop}%
\bibitem [{\citenamefont {Huang}\ \emph {et~al.}(2020)\citenamefont {Huang},
  \citenamefont {Qu}, \citenamefont {Chuang}, \citenamefont {Chiang},
  \citenamefont {Lin},\ and\ \citenamefont {Chien}}]{Huang}%
  \BibitemOpen
  \bibfield  {author} {\bibinfo {author} {\bibfnamefont {S.~Y.}\ \bibnamefont
  {Huang}}, \bibinfo {author} {\bibfnamefont {D.}~\bibnamefont {Qu}}, \bibinfo
  {author} {\bibfnamefont {T.~C.}\ \bibnamefont {Chuang}}, \bibinfo {author}
  {\bibfnamefont {C.~C.}\ \bibnamefont {Chiang}}, \bibinfo {author}
  {\bibfnamefont {W.}~\bibnamefont {Lin}}, \ and\ \bibinfo {author}
  {\bibfnamefont {C.~L.}\ \bibnamefont {Chien}},\ }\href {\doibase
  10.1063/5.0032368} {\bibfield  {journal} {\bibinfo  {journal} {Appl. Phys.
  Lett.}\ }\textbf {\bibinfo {volume} {117}},\ \bibinfo {pages} {190501}
  (\bibinfo {year} {2020})}\BibitemShut {NoStop}%
\bibitem [{\citenamefont {Lin}\ and\ \citenamefont {Chien}(2018)}]{Lin}%
  \BibitemOpen
  \bibfield  {author} {\bibinfo {author} {\bibfnamefont {W.}~\bibnamefont
  {Lin}}\ and\ \bibinfo {author} {\bibfnamefont {C.~L.}\ \bibnamefont
  {Chien}},\ }\href@noop {} {\bibfield  {journal} {\bibinfo  {journal}
  {arXiv:1804.01392}\ } (\bibinfo {year} {2018})}\BibitemShut {NoStop}%
\bibitem [{\citenamefont {Shan}\ \emph {et~al.}(2015)\citenamefont {Shan},
  \citenamefont {Zhou},\ and\ \citenamefont {Xiao}}]{ShanW}%
  \BibitemOpen
  \bibfield  {author} {\bibinfo {author} {\bibfnamefont {W.-Y.}\ \bibnamefont
  {Shan}}, \bibinfo {author} {\bibfnamefont {J.}~\bibnamefont {Zhou}}, \ and\
  \bibinfo {author} {\bibfnamefont {D.}~\bibnamefont {Xiao}},\ }\href {\doibase
  10.1103/physrevb.91.035402} {\bibfield  {journal} {\bibinfo  {journal} {Phys.
  Rev. B}\ }\textbf {\bibinfo {volume} {91}},\ \bibinfo {pages} {035402}
  (\bibinfo {year} {2015})}\BibitemShut {NoStop}%
\bibitem [{\citenamefont {S{\l}awi{\'{n}}ska}\ \emph
  {et~al.}(2019)\citenamefont {S{\l}awi{\'{n}}ska}, \citenamefont {Cerasoli},
  \citenamefont {Wang}, \citenamefont {Postorino}, \citenamefont {Supka},
  \citenamefont {Curtarolo}, \citenamefont {Fornari},\ and\ \citenamefont
  {Nardelli}}]{Slawinska}%
  \BibitemOpen
  \bibfield  {author} {\bibinfo {author} {\bibfnamefont {J.}~\bibnamefont
  {S{\l}awi{\'{n}}ska}}, \bibinfo {author} {\bibfnamefont {F.~T.}\ \bibnamefont
  {Cerasoli}}, \bibinfo {author} {\bibfnamefont {H.}~\bibnamefont {Wang}},
  \bibinfo {author} {\bibfnamefont {S.}~\bibnamefont {Postorino}}, \bibinfo
  {author} {\bibfnamefont {A.}~\bibnamefont {Supka}}, \bibinfo {author}
  {\bibfnamefont {S.}~\bibnamefont {Curtarolo}}, \bibinfo {author}
  {\bibfnamefont {M.}~\bibnamefont {Fornari}}, \ and\ \bibinfo {author}
  {\bibfnamefont {M.~B.}\ \bibnamefont {Nardelli}},\ }\href {\doibase
  10.1088/2053-1583/ab0146} {\bibfield  {journal} {\bibinfo  {journal} {2D
  Mater.}\ }\textbf {\bibinfo {volume} {6}},\ \bibinfo {pages} {025012}
  (\bibinfo {year} {2019})}\BibitemShut {NoStop}%
\bibitem [{\citenamefont {Yao}\ and\ \citenamefont {Fang}(2005)}]{Yao}%
  \BibitemOpen
  \bibfield  {author} {\bibinfo {author} {\bibfnamefont {Y.}~\bibnamefont
  {Yao}}\ and\ \bibinfo {author} {\bibfnamefont {Z.}~\bibnamefont {Fang}},\
  }\href {\doibase 10.1103/physrevlett.95.156601} {\bibfield  {journal}
  {\bibinfo  {journal} {Phys. Rev. Lett.}\ }\textbf {\bibinfo {volume} {95}},\
  \bibinfo {pages} {156601} (\bibinfo {year} {2005})}\BibitemShut {NoStop}%
\bibitem [{\citenamefont {Da}\ \emph {et~al.}(2020)\citenamefont {Da},
  \citenamefont {Song}, \citenamefont {Dong}, \citenamefont {Ye},\ and\
  \citenamefont {Yan}}]{Da}%
  \BibitemOpen
  \bibfield  {author} {\bibinfo {author} {\bibfnamefont {H.}~\bibnamefont
  {Da}}, \bibinfo {author} {\bibfnamefont {Q.}~\bibnamefont {Song}}, \bibinfo
  {author} {\bibfnamefont {P.}~\bibnamefont {Dong}}, \bibinfo {author}
  {\bibfnamefont {H.}~\bibnamefont {Ye}}, \ and\ \bibinfo {author}
  {\bibfnamefont {X.}~\bibnamefont {Yan}},\ }\href {\doibase 10.1063/1.5118327}
  {\bibfield  {journal} {\bibinfo  {journal} {J. Appl. Phys.}\ }\textbf
  {\bibinfo {volume} {127}},\ \bibinfo {pages} {023903} (\bibinfo {year}
  {2020})}\BibitemShut {NoStop}%
\bibitem [{\citenamefont {Gradhand}\ \emph {et~al.}(2012)\citenamefont
  {Gradhand}, \citenamefont {Fedorov}, \citenamefont {Pientka}, \citenamefont
  {Zahn}, \citenamefont {Mertig},\ and\ \citenamefont {Györffy}}]{Grad}%
  \BibitemOpen
  \bibfield  {author} {\bibinfo {author} {\bibfnamefont {M.}~\bibnamefont
  {Gradhand}}, \bibinfo {author} {\bibfnamefont {D.~V.}\ \bibnamefont
  {Fedorov}}, \bibinfo {author} {\bibfnamefont {F.}~\bibnamefont {Pientka}},
  \bibinfo {author} {\bibfnamefont {P.}~\bibnamefont {Zahn}}, \bibinfo {author}
  {\bibfnamefont {I.}~\bibnamefont {Mertig}}, \ and\ \bibinfo {author}
  {\bibfnamefont {B.~L.}\ \bibnamefont {Györffy}},\ }\href {\doibase
  10.1088/0953-8984/24/21/213202} {\bibfield  {journal} {\bibinfo  {journal}
  {J. Phys. Condens. Matter}\ }\textbf {\bibinfo {volume} {24}},\ \bibinfo
  {pages} {213202} (\bibinfo {year} {2012})}\BibitemShut {NoStop}%
\bibitem [{\citenamefont {Qu}\ \emph {et~al.}(2019)\citenamefont {Qu},
  \citenamefont {Nakamura},\ and\ \citenamefont {Hayashi}}]{Qu}%
  \BibitemOpen
  \bibfield  {author} {\bibinfo {author} {\bibfnamefont {G.}~\bibnamefont
  {Qu}}, \bibinfo {author} {\bibfnamefont {K.}~\bibnamefont {Nakamura}}, \ and\
  \bibinfo {author} {\bibfnamefont {M.}~\bibnamefont {Hayashi}},\ }\href@noop
  {} {\bibfield  {journal} {\bibinfo  {journal} {arXiv:1901.05651}\ } (\bibinfo
  {year} {2019})}\BibitemShut {NoStop}%
\bibitem [{\citenamefont {Kim}\ \emph {et~al.}(2019)\citenamefont {Kim},
  \citenamefont {Kim}, \citenamefont {Shin}, \citenamefont {Lee}, \citenamefont
  {Sinova}, \citenamefont {Park},\ and\ \citenamefont {Jin}}]{Ki}%
  \BibitemOpen
  \bibfield  {author} {\bibinfo {author} {\bibfnamefont {J.}~\bibnamefont
  {Kim}}, \bibinfo {author} {\bibfnamefont {K.-W.}\ \bibnamefont {Kim}},
  \bibinfo {author} {\bibfnamefont {D.}~\bibnamefont {Shin}}, \bibinfo {author}
  {\bibfnamefont {S.-H.}\ \bibnamefont {Lee}}, \bibinfo {author} {\bibfnamefont
  {J.}~\bibnamefont {Sinova}}, \bibinfo {author} {\bibfnamefont
  {N.}~\bibnamefont {Park}}, \ and\ \bibinfo {author} {\bibfnamefont
  {H.}~\bibnamefont {Jin}},\ }\href {\doibase 10.1038/s41467-019-11964-6}
  {\bibfield  {journal} {\bibinfo  {journal} {Nat. Commun.}\ }\textbf {\bibinfo
  {volume} {10}},\ \bibinfo {pages} {3965} (\bibinfo {year}
  {2019})}\BibitemShut {NoStop}%
\bibitem [{\citenamefont {Guo}\ \emph {et~al.}(2008)\citenamefont {Guo},
  \citenamefont {Murakami}, \citenamefont {Chen},\ and\ \citenamefont
  {Nagaosa}}]{GuoG}%
  \BibitemOpen
  \bibfield  {author} {\bibinfo {author} {\bibfnamefont {G.~Y.}\ \bibnamefont
  {Guo}}, \bibinfo {author} {\bibfnamefont {S.}~\bibnamefont {Murakami}},
  \bibinfo {author} {\bibfnamefont {T.-W.}\ \bibnamefont {Chen}}, \ and\
  \bibinfo {author} {\bibfnamefont {N.}~\bibnamefont {Nagaosa}},\ }\href
  {\doibase 10.1103/physrevlett.100.096401} {\bibfield  {journal} {\bibinfo
  {journal} {Phys. Rev. Lett.}\ }\textbf {\bibinfo {volume} {100}},\ \bibinfo
  {pages} {096401} (\bibinfo {year} {2008})}\BibitemShut {NoStop}%
\bibitem [{\citenamefont {Zhou}\ \emph {et~al.}(2019)\citenamefont {Zhou},
  \citenamefont {Qiao}, \citenamefont {Bournel},\ and\ \citenamefont
  {Zhao}}]{Zh}%
  \BibitemOpen
  \bibfield  {author} {\bibinfo {author} {\bibfnamefont {J.}~\bibnamefont
  {Zhou}}, \bibinfo {author} {\bibfnamefont {J.}~\bibnamefont {Qiao}}, \bibinfo
  {author} {\bibfnamefont {A.}~\bibnamefont {Bournel}}, \ and\ \bibinfo
  {author} {\bibfnamefont {W.}~\bibnamefont {Zhao}},\ }\href {\doibase
  10.1103/physrevb.99.060408} {\bibfield  {journal} {\bibinfo  {journal} {Phys.
  Rev. B}\ }\textbf {\bibinfo {volume} {99}},\ \bibinfo {pages} {060408(R)}
  (\bibinfo {year} {2019})}\BibitemShut {NoStop}%
\bibitem [{\citenamefont {Feng}\ \emph {et~al.}(2012)\citenamefont {Feng},
  \citenamefont {Yao}, \citenamefont {Zhu}, \citenamefont {Zhou}, \citenamefont
  {Yao},\ and\ \citenamefont {Xiao}}]{Feng}%
  \BibitemOpen
  \bibfield  {author} {\bibinfo {author} {\bibfnamefont {W.}~\bibnamefont
  {Feng}}, \bibinfo {author} {\bibfnamefont {Y.}~\bibnamefont {Yao}}, \bibinfo
  {author} {\bibfnamefont {W.}~\bibnamefont {Zhu}}, \bibinfo {author}
  {\bibfnamefont {J.}~\bibnamefont {Zhou}}, \bibinfo {author} {\bibfnamefont
  {W.}~\bibnamefont {Yao}}, \ and\ \bibinfo {author} {\bibfnamefont
  {D.}~\bibnamefont {Xiao}},\ }\href {\doibase 10.1103/physrevb.86.165108}
  {\bibfield  {journal} {\bibinfo  {journal} {Phys. Rev. B}\ }\textbf {\bibinfo
  {volume} {86}},\ \bibinfo {pages} {165108} (\bibinfo {year}
  {2012})}\BibitemShut {NoStop}%
\bibitem [{\citenamefont {Xiao}\ \emph {et~al.}(2010)\citenamefont {Xiao},
  \citenamefont {Chang},\ and\ \citenamefont {Niu}}]{Xiao}%
  \BibitemOpen
  \bibfield  {author} {\bibinfo {author} {\bibfnamefont {D.}~\bibnamefont
  {Xiao}}, \bibinfo {author} {\bibfnamefont {M.-C.}\ \bibnamefont {Chang}}, \
  and\ \bibinfo {author} {\bibfnamefont {Q.}~\bibnamefont {Niu}},\ }\href
  {\doibase 10.1103/revmodphys.82.1959} {\bibfield  {journal} {\bibinfo
  {journal} {Rev. Mod. Phys.}\ }\textbf {\bibinfo {volume} {82}},\ \bibinfo
  {pages} {1959} (\bibinfo {year} {2010})}\BibitemShut {NoStop}%
\bibitem [{\citenamefont {Wang}\ \emph {et~al.}(2014)\citenamefont {Wang},
  \citenamefont {Deorani}, \citenamefont {Qiu}, \citenamefont {Kwon},\ and\
  \citenamefont {Yang}}]{WangY}%
  \BibitemOpen
  \bibfield  {author} {\bibinfo {author} {\bibfnamefont {Y.}~\bibnamefont
  {Wang}}, \bibinfo {author} {\bibfnamefont {P.}~\bibnamefont {Deorani}},
  \bibinfo {author} {\bibfnamefont {X.}~\bibnamefont {Qiu}}, \bibinfo {author}
  {\bibfnamefont {J.~H.}\ \bibnamefont {Kwon}}, \ and\ \bibinfo {author}
  {\bibfnamefont {H.}~\bibnamefont {Yang}},\ }\href {\doibase
  10.1063/1.4898593} {\bibfield  {journal} {\bibinfo  {journal} {Appl. Phys.
  Lett.}\ }\textbf {\bibinfo {volume} {105}},\ \bibinfo {pages} {152412}
  (\bibinfo {year} {2014})}\BibitemShut {NoStop}%
\bibitem [{\citenamefont {Zhang}\ \emph {et~al.}(2016)\citenamefont {Zhang},
  \citenamefont {Han}, \citenamefont {Yang}, \citenamefont {Sun}, \citenamefont
  {Zhang}, \citenamefont {Yan},\ and\ \citenamefont {Parkin}}]{ZhanW}%
  \BibitemOpen
  \bibfield  {author} {\bibinfo {author} {\bibfnamefont {W.}~\bibnamefont
  {Zhang}}, \bibinfo {author} {\bibfnamefont {W.}~\bibnamefont {Han}}, \bibinfo
  {author} {\bibfnamefont {S.-H.}\ \bibnamefont {Yang}}, \bibinfo {author}
  {\bibfnamefont {Y.}~\bibnamefont {Sun}}, \bibinfo {author} {\bibfnamefont
  {Y.}~\bibnamefont {Zhang}}, \bibinfo {author} {\bibfnamefont
  {B.}~\bibnamefont {Yan}}, \ and\ \bibinfo {author} {\bibfnamefont {S.~S.~P.}\
  \bibnamefont {Parkin}},\ }\href {\doibase 10.1126/sciadv.1600759} {\bibfield
  {journal} {\bibinfo  {journal} {Sci. Adv.}\ }\textbf {\bibinfo {volume}
  {2}},\ \bibinfo {pages} {e1600759} (\bibinfo {year} {2016})}\BibitemShut
  {NoStop}%
\bibitem [{\citenamefont {Kubo}(1957)}]{Kubo}%
  \BibitemOpen
  \bibfield  {author} {\bibinfo {author} {\bibfnamefont {R.}~\bibnamefont
  {Kubo}},\ }\href {\doibase 10.1143/jpsj.12.570} {\bibfield  {journal}
  {\bibinfo  {journal} {J. Phys. Soc. Japan}\ }\textbf {\bibinfo {volume}
  {12}},\ \bibinfo {pages} {570} (\bibinfo {year} {1957})}\BibitemShut
  {NoStop}%
\bibitem [{\citenamefont {Saberi-Pouya}\ \emph {et~al.}(2017)\citenamefont
  {Saberi-Pouya}, \citenamefont {Vazifehshenas}, \citenamefont {Salavati-fard},
  \citenamefont {Farmanbar},\ and\ \citenamefont {Peeters}}]{SPS}%
  \BibitemOpen
  \bibfield  {author} {\bibinfo {author} {\bibfnamefont {S.}~\bibnamefont
  {Saberi-Pouya}}, \bibinfo {author} {\bibfnamefont {T.}~\bibnamefont
  {Vazifehshenas}}, \bibinfo {author} {\bibfnamefont {T.}~\bibnamefont
  {Salavati-fard}}, \bibinfo {author} {\bibfnamefont {M.}~\bibnamefont
  {Farmanbar}}, \ and\ \bibinfo {author} {\bibfnamefont {F.~M.}\ \bibnamefont
  {Peeters}},\ }\href {\doibase 10.1103/PhysRevB.96.075411} {\bibfield
  {journal} {\bibinfo  {journal} {Phys. Rev. B}\ }\textbf {\bibinfo {volume}
  {96}},\ \bibinfo {pages} {075411} (\bibinfo {year} {2017})}\BibitemShut
  {NoStop}%
\end{thebibliography}%
\end{document}